\documentclass[aps,prl,twocolumn,float]{revtex4}
\usepackage{amsmath,bm,epsfig}
\newcommand{\B}[1]{{\bm{#1}}}%% Bold Roman & Greek Lower & Upper Case
\usepackage[dvips]{color}
\begin{document}
\title{Determining the Inter-Particle Force-Laws in Amorphous Solids from a Visual Image}
\author{Oleg Gendelman$^{1,2}$, Yoav G. Pollack$^{1}$ and Itamar Procaccia$^{1}$}
\affiliation{$^{1}$Dept. of Chemical Physics, The Weizmann Institute of Science,  Rehovot 76100, Israel\\
Faculty of Mechanical Engineering, Technion, Haifa 32000, Israel}
\begin{abstract}
We consider the problem of how to determine the force laws in an amorphous system of interacting particles. Given the positions of the centers of mass of the constituent particles we propose a new algorithm to determine the inter-particle force-laws. Having $n$ different types of constituents we determine
the coefficients in the Laurent polynomials for the $n(n+1)/2$ possibly different force-laws. A visual providing the particle positions in addition to a measurement of the pressure is all that is required. The algorithm proposed includes
a part that can correct for experimental errors in the positions of the particles. Such a correction of unavoidable
measurement errors is expected to benefit many experiments in the field.
\end{abstract}
\maketitle

The impressive technological progress that allows an accurate determination of the positions of particles
in two and three dimensional amorphous systems \cite{96CG,09GK,00WCLSW,07PSW,11CMIC} opens up new possibilities for improving the understanding of these versatile
and fascinating materials. For example in colloidal systems, microscopic information at the single-particle level is obtained with a laser scanning confocal microscope \cite{04SCWS}. In this Letter we propose a new method to determine the force-laws governing inter-particle forces, based on accurate visualizations of amorphous systems in which the provided information is the positions of the centers of mass of all the involved particles and the global pressure of the system. We will show
that this information is sufficient for an accurate determination of the inter-particle force-laws even when the system contains particles of different types \cite{04HG,06ZLWD,14HMRA}. The method does not require an explicit knowledge of
external forces, these are determined as well by the proposed algorithm. The present method is applicable in fact
to any type of amorphous material as long as the forces are central and frictional forces are absent. For systems with frictional forces one needs a different approach, cf. \cite{16GPPSZ}.

At zero temperature the position of every particle in a mechanically stable system is fixed. Not so in thermal
systems where particles suffer temperature fluctuations. For the purposes of the present discussion we assume that
one can determine the average position of each particle by taking sufficiently long time averages, but shorter than
typical diffusion times during which particles can escape out of their local cages. We will denote the average positions of $N$ particles in a volume $V$ as $\{\B r_i\}_{i=1}^N$. For a frictionless system in mechanical equilibrium we assert
that the inter-particle forces are central, directed along the inter-particle vector distance $\B r_{ij}\equiv \B r_j-\B r_i$. The inter-particle forces are assumed to depend in an a-priori unknown way on the scalar distance $r_{ij}$, i.e.
$\B f_{ij} = \B f_{ij} (r_{ij})$. We allow different types of particles interacting via different force
laws $\B f_{ij}^{AB}$ where the notation $A, B$ runs over the different species. No knowledge of the external
forces is required. We assume that the global pressure is known. Finally, we assume that the inter-particle forces vanish sufficiently rapidly when $r_{ij}$ exceed a few
particle distances. At this point we exclude 3-body and higher order interactions.

In typical physical systems particles cannot be brought infinitely close to each other, meaning that the forces between
them become repulsive and very large at some inter-particle distance $r_{ij} \ge 0$. We therefore
acknowledge below the possible existence of a singularity in the force-laws, but insist that (i) the singularity does not
have to be at $r_{ij}=0$, and (ii) the position of the singularity may differ for each interacting pair of species.
It is quite remarkable, as we show below, that it is not necessary to know the position of the singularity a-priori.

\begin{figure}
\includegraphics[scale = 0.30]{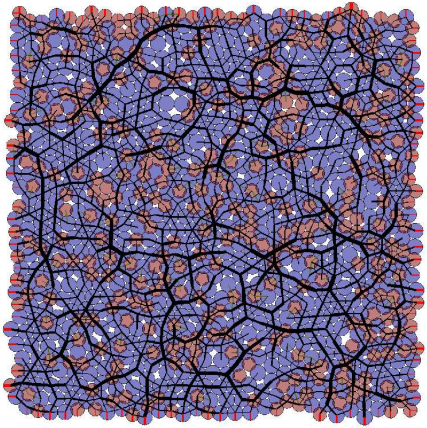}
\caption{An example of a typical configuration of an amorphous solid made of 1000 particles of two diameters. The external forces are those between the walls
and the particles adjacent to the walls, marked in red short lines. The force chains, made from the 20\% strongest forces, are indicated by black lines between particles whose width is proportional to the strength of the inter-particle forces. The particle ``diameters" used in this figure correspond to the distance at which the $AA$ and $BB$ interactions vanish. The $AB$ interaction force vanish at shorter distances so that a lot of $AB$ pairs appear more compressed than they really are.}
\label{image}
\end{figure}

To exemplify the new algorithm we consider a 2-dimensional system of $N$ particles with
$c$ binary contacts enclosed in a rectangular box . In 2-dimensions the starting point of the algorithm is furnished by the mechanical equilibrium constraints:
\begin{equation}
\B M|F\rangle=0 \ ,
\label{MF}
\end{equation}
where $|F\rangle$ is a vector of the magnitudes of the inter-particle forces, followed by the x and y components of the external forces.
\begin{equation}
|F\rangle =\left(
\begin{array}{c}
f_{ij} \\
f_i^{x,ext} \\
f_i^{y,ext} \\
\end{array}
\right).\label{force}
\end{equation}
The external forces are assigned to particles that are close to the boundaries; particles that are near the west or east walls contribute an $x$ component entry, whereas a $y$ component entry is contributed by particles near the north and south walls, cf. Fig.~\ref{image}.
Only particles that are stuck in corners can have both $x$ and $y$ entries. All the other possible external forces are assumed to vanish and are {\em not} included in this vector; gravity can be added with impunity but at present we disregard it. It is important to stress that the algorithm proposed below does NOT require a measurement of the external forces, they are a part of the result of the calculation. Since there are $c$ contact forces and $e$ external force entries, the length of the vector $|F\rangle$ is ($c+e$).

The matrix $\B M$ in Eq.~(\ref{MF}) specifies the directions of the force vectors. Denote the unit vector
 in the direction of the vector distance between the centers of mass of particles $i$ and $j$ by $\B {\hat n}_{ij}$. Then the entries of $\B M$ display the projections $\hat n^x_{ij}$ and $\hat n^y_{ij}$ as appropriate. In addition, since the external forces are already given in terms of $x$ and $y$ components, their entries in $\B M$ are simply 1 or -1.

The analysis below requires a subdivision of $\B M$ into two parts $\B M_1$ and $\B M_2$.
 \begin{equation}
 \B M=\left(
 \begin{array}{ccc}
 \B M_1&, & \B M_2\\
 \end{array}
 \right) \ ,
 \end{equation}
where $\B M_1$ is the $2N \times c$ matrix that accounts for the inter-particle forces and $\B M_2$ is a $2N \times e$ matrix that accounts for the external forces.

Our aim here is to employ the mechanical constraints to determine the force laws. To this aim
the inter-particle force magnitudes are presented as Laurent polynomials:
\begin{align}
f^{AB}_{ij}=\overset{\ell_2}{\sum _{k=\ell_1}}a_k^{AB}\left(r_{ij}-r_0^{AB}\right)^k \ ,
\label{eq:Taylor}
\end{align}
where $\ell_1$, $\ell_2$ are the most negative and most positive powers in the expansion respectively. Below we will
denote the number of terms in the expansion as $\ell\equiv \ell_2-\ell_1+1$. $AB$ denotes the interaction type. For example in the case of a binary system these will be  (AA,BB,AB), as  determined by the nature of the particles $i,j$. $r_0^{AB}$ are the positions of the possible singularities around which we expand the forces for each type of interaction.
The coefficients $a_k^{AB}$ can be grouped into a vector $|a\rangle$ of size $n(n+1)\ell/2 $. For a binary
system its transpose reads
\begin{equation}
\langle a| =\left(
a_{\ell_1}^{AA} \cdots
a_{\ell_2}^{AA}
a_{\ell_1}^{BB} \cdots
a_{\ell_2}^{BB}
a_{\ell_1}^{AB}\cdots
a_{\ell_2}^{AB}
\right) \ ,
\end{equation}
and the force vector can now be written as:
\begin{equation}
|F\rangle =\left(
\begin{array}{c}
\B S |a\rangle \\
f_i^{x,ext} \\
f_i^{y,ext} \\
\end{array}
\right) \ ,
\end{equation}
where $\B S$ is the appropriate $c\times n(n+1)\ell/2$ matrix containing the Laurent monomials. An example of some of the components of the $\B S$
matrix for a minimal (unrealistic) expansion with $\ell_1=-1$ and $\ell_2=1$ in a binary system will read
\begin{widetext}
\begin{equation}
 \B S=
{
\left(
\begin{array}{cccccccccc}
\left(r_1-r_0^{AA}\right)^{-1} & 1 & \left(r_1-r_0^{AA}\right) & 0 & 0 & 0 & 0 & 0 & 0 &   \\
0 & 0 & 0 & 0 & 0 & 0 & \left(r_2-r_0^{AB}\right)^{-1} & 1 & \left(r_2-r_0^{AB}\right) &  \\
0 & 0 & 0 & \left(r_3-r_0^{BB}\right)^{-1} & 1 & \left(r_3-r_0^{BB}\right) & 0 & 0 & 0 &   \\
 &   &   &  & : &   &   &   &   &   \\
\end{array}
\right) \ .
}
\end{equation}
\end{widetext}

A solution of the coefficients of the Laurent expansion is not unique without fixing one
scale parameter. A natural choice for such a parameter is the pressure in the system.
\begin{equation}
P=\frac{1}{2}\left(\sigma _{xx}+\sigma _{yy}\right)=\frac{1}{4}\left(\frac{1}{L_x}\sum _if_i^{x,ext}+\frac{1}{L_y}\sum _if_i^{y,ext}\right) \ ,
\end{equation}
where the extra half factor comes from the fact that the summation is on all the forces, instead of just one side of the box.
We can now add the equation of the pressure to the force balance constraints to get:
\begin{equation}
\left(
\begin{array}{cc}
\B M_1 & \B M_2 \\
0 & \frac{1}{4L_x}\frac{1}{4L_x}...\frac{1}{4L_y} \\
\end{array}
\right)
\left(
\begin{array}{c}
f_{ij} \\
f_i^{x,ext} \\
f_i^{y,ext} \\
\end{array}
\right)
=\left(
\begin{array}{c}
0 \\
: \\
P \\
\end{array}
\right) \ .
\end{equation}
This can be converted to a form that conveniently groups all the unknowns into one vector $|u\rangle$:
\begin{align}
\left(
\begin{array}{cc}
\B M_1 \B S & \B M_2 \\
0 & \frac{1}{4L_x}\frac{1}{4L_x}...\frac{1}{4L_y} \\
\end{array}
\right)
\left(
\begin{array}{c}
|a\rangle \\
f_i^{x,ext} \\
f_i^{y,ext} \\
\end{array}
\right)
\equiv \B Y|u\rangle
=\left(
\begin{array}{c}
0 \\
: \\
P \\
\end{array}
\right) \ ,
\label{pre}
\end{align}
where $\B Y$ is a matrix of size $(2N+1) \times (\ell \cdot n(n+1)/2+e)$.
We now multiply by $\B Y^T$ from the left
\begin{equation}
\B Y^T\B Y|u\rangle
=\B Y^T\left(
\begin{array}{c}
0 \\
: \\
P \\
\end{array}
\right) \ .
\label{ls}
\end{equation}
Since $2N+1\gg \ell \cdot n(n+1)/2+e$ this equation should be well-posed. We can therefore invert
with impunity to get
\begin{align}
|u\rangle \equiv
 \left(
\begin{array}{c}
|a\rangle \\
f_i^{x,ext} \\
f_i^{y,ext} \\
\end{array}
\right)
=\left(\B Y^T\B Y\right)^{-1}\B Y^T\left(
\begin{array}{c}
0 \\
: \\
P \\
\end{array}
\right) \ .
\label{solution}
\end{align}
In fact, since the Laurent expansion is finite, the solution that we seek cannot be exact. Therefore the analytic
inversion Eq.~(\ref{solution}) should be understood as a least-squares solution for
the coefficients of Laurent polynomials and the external forces.  In practice this is achieved by using the {\em mldivide} function in Matlab.
We note in passing that it is not guaranteed that the matrix $\B Y^T \B Y$ has only  nonzero eigenvalues.
Nevertheless, even when it has zero eigenvalues, the RHS of Eq.~(\ref{ls}) is orthogonal to the eigenfunctions associated
with these eigenvalues, and the inversion is still possible. A word of caution: if we try to over-fit and increase the number of expansion coefficients and/or the number of external forces such that $2N+1< \ell \cdot n(n+1)/2+e$ one may eventually run into trouble,
since the zero modes of the matrix $\B Y^T \B Y$ may cease being orthogonal to the RHS of Eq.~(\ref{ls}). This discussion
can be clarified by presenting the solution for $|u\rangle$ as an expansion in the eigenfunctions
$\Psi_i$ of $Y^T Y$:
\begin{equation}
|u\rangle =\sum_i \frac{\langle \Psi_i|Y^T|t\rangle}{\lambda_i }|\Psi_i\rangle \ .
\label{eq:eigenmodes}
\end{equation}
where $\lambda_i$ are the eigenvalues and
\begin{equation}
|t\rangle\equiv \left(
\begin{array}{c}
0 \\
: \\
P \\
\end{array}
\right) \ .
\end{equation}
We see explicitly that eigenfunctions that are orthogonal to the vector $Y^T |t\rangle$ do not contribute to the solution $|u\rangle$. As long as the zero modes fulfill this orthogonality condition, they pose no problem. Such zero modes can arise from voids, and from ``rattlers", i.e. particles within the voids, which do not contribute to the pressure of the system. At this point we comment that an extension of the present approach to 3-dimensions amounts to adding external force in the
$z$ direction and making the unit vectors in the matrix $\B M$ 3-dimensional with an obvious size adjustment of the matrix.

In the rest of this Letter we exemplify the efficacy of the method by considering a 2-dimensional typical glass-former,
i.e. a Kob-Andersen model \cite{95KA}.  The model employs two types of particles A and B interacting via Lennard-Jones forces. Thus $n=2$ and we have three types of interactions,  AA, BB and AB. The simulation is done with $N=1000$ particles in a square box as presented in Fig.~\ref{image}. The walls of the box exert Hookean restoring forces on the particles that attempt to cross them. By pushing the walls inwardly we can  compress the system from an initial low density configuration to any desired density. This compression is done quasistatically at zero temperature by performing a conjugate gradient energy minimization after every infinitesimal step of compression.
In the simulations described below we employed a Hookean force constant of magnitude 100 in Lennard-Jones units.

The smoothed Lennard-Jones potentials were chosen to be
\begin{align}
U({\B r_{ij}}) =
\left\{\begin{array}{ccc}
4\epsilon_{ij} \Big[ \big(\frac{\sigma_{ij}}{r_{ij}}\big)^{12}  - \big(\frac{\sigma_{ij}}{r_{ij}}\big)^{6}+C_0 & \\
+C_2\big(\frac{\sigma_{ij}}{r_{ij}}\big)^{-2}
+C_4\big(\frac{\sigma_{ij}}{r_{ij}}\big)^{-4}\Big]&, & \text{for } \frac{r_{ij}}{\sigma_{ij}} \leq x_c \\
0&, & \text{for } \frac{r_{ij}}{\sigma_{ij}} > x_c \ .
\end{array}\right.
\end{align}
Here $r_{ij}\equiv|\B r_{ij}|$. The coefficients $C_0$,$C_2$,$C_4$ are chosen in such a way that the potential and
its first and second derivatives vanish at the cutoff $x_c = 2.5$.

 In the present variant of the model, 65\% of the particles are of type $A$ and 35\% of type $B$, with particle `diameters' and interaction energy scales defined by $\sigma_{AA} = 1$, $\sigma_{AB} = 0.8$ and $\sigma_{BB} = 0.88$, and $\epsilon_{AA} = 1$, $\epsilon_{AB} = 1.5$ and $\epsilon_{BB} = 0.5$, respectively. Lengths and energies are henceforth given in terms of $\sigma_{AA}$ and $\epsilon_{AA}$, while time units are given by $\sqrt{m \sigma^2_{AA}/\epsilon_{AA}}$. Both the Boltzmann constant $k_B$ and the mass of the particles are taken to be unity.

Obviously, these forces become singular at $r_0^{AB}=0$. Nevertheless it turns out that the Laurent expansion that
we use allows substantial freedom. First,
the powers used in the Laurent polynomial could be changed in a wide range without major changes in the results with $-19\le \ell_1 \le -3$ and $3\le \ell_2 \le 13$. Second, we find that as long as we set $r^{AA}_0\le 0.88$, $r^{BB}_{0}\le 0.74$ and $r^{AB}_{0}\le 0.7$ the final actual results were almost invariant.

\begin{figure}[h!]
\vskip 0.5 cm
\hskip -0.4 cm
\includegraphics[scale = 0.19]{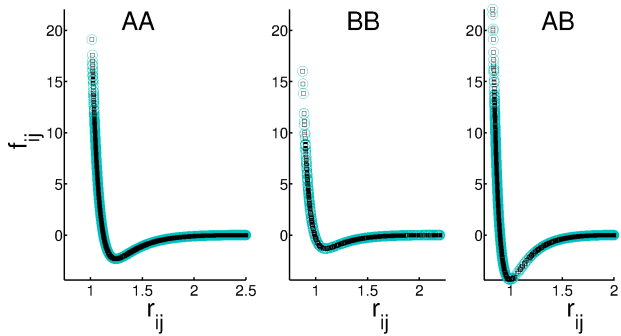}
\caption{Comparison between the predicted (scalar) forces of interaction (green circles), and the ones used in the simulation
(black squares). In the present
comparison the exact particle positions are provided.}
\label{comparison}
\end{figure}

A typical comparison between the exact Lennard-Jones forces and their Laurent approximants as obtained from
this algorithm are shown in Fig.~\ref{comparison}.
%%%%%%%%%%%%%%%%%%%%%%%%%%%%%%%%%%%%%%%%%%%%%%%%%%%%%%%%%%%%%%%%%%%%%%%%

Of course, the excellent agreement seen in Fig.~\ref{comparison} stems in part from the fact that the particle
positions were provided with machine precision. In reality, an experimental visual of a real system will
contain errors in the particle positions. It is useful therefore to assess the efficacy of the present
algorithm in situations where there exists a realistic error in the particle positions, and if possible
to offer a way to correct for such errors.

\begin{figure}
\includegraphics[scale = 0.21]{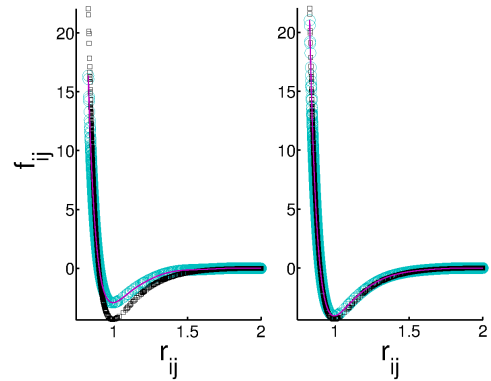}
\caption{An example of the effect of errors in particle positions and the correction. Left panel: the AB force as predicted from the erroneous data. Right panel: the same force as predicted after the Initial errors in the particle positions are corrected with the help of the pseudo gradient-descent algorithm.}
\label{compare2}
\end{figure}

To begin with, consider in Fig.~\ref{compare2}, the predicted AB force
law for the very same system shown
in Fig.~\ref{comparison} but with the positions of the particles perturbed by a random jitter from a normal distribution
with standard deviation of
$10^{-4}$ in Lennard-Jones units. The comparison now appears poor, with considerable
deviations between the input Lennard-Jones force and its prediction. Similar errors appear in the other forces. We must therefore come up with a method to correct for these discrepancies.
%%%%%%%%%%%%%%%%%%%%%%%%%%%%%%%%%%%%%%%%%%%%%%%%%%%%%%%%%%%%%%%%%%
%%%%%%%%%%%%%%%%%%%%%%%%%%%%%%%%%%%%%%%%%%%%%%%%%%%%%%%%%%%%%%%%
The procedure to correct the errors in particle position is an iterative process composed of three steps: In the first step
we compute the force laws as shown in Fig.~\ref{compare2} from the erroneous position data. Secondly we use
these force laws to compute the net force on each particle. Due to the errors in force laws, the net forces
are not annulled, and therefore we can execute the third step, which is a pseudo gradient-descent step where each particle is displaced in the direction of the net force predicted for it. The amount of displacement of each particle is chosen as the magnitude of the calculated net force times the learning rate $\alpha$. The learning rate is chosen somewhat arbitrarily as always in gradient-descent. It should be chosen to have the largest value that still leads to convergence of the procedure.
The third step is a recalculation of the coefficients of the Laurent expansion as detailed above.

The procedure converges, with the force laws obtained as shown in Fig.~\ref{compare2} in the right panel for the AB
interaction. In fact we could increase the initial error in positions by an order of magnitude and still the procedure
converged.

In summary, we have presented a simple and practical method to determine the force laws in amorphous systems of
particles whose center-of-mass positions (or its average over time) are known. When the data is precise, the
force laws are determined to high accuracy. When the data is noisy, we indicated how one can correct for the
errors in particle positions by implementing an iterative procedure in which the ``wrong" forces are used to correct
for the positions of the particles. This results in more accurate force-laws but also with an improved knowledge of the
correct particle positions. This is a very simple scheme, and it can be improved. For example one
can use more than one realization for the same system to improve even further the predicted force laws.
This and further improvements of the method will be discussed in a follow-up publication.

This work had been supported in part by an ERC ``ideas" grant STANPAS and by the Minerva Foundation,
Munich, Germany.

\end{document}